\documentclass[12pt]{article}
\usepackage{scicite}
\usepackage{times}
\usepackage{graphicx}
\usepackage{physics}
\usepackage{xcolor}%
\usepackage{float}

\newenvironment{sciabstract}{%
\begin{quote} \bf}
{\end{quote}}

\newcounter{lastnote}

\title{Optically-detected room-temperature spin-coherence from the GaV$^-$ in diamond}

\author
{	Adam D. Dodson$^{1}$,
	Tiamike Dudley$^{1,2}$,
    Luca Basso$^{1}$,
    Leopoldo Diaz$^{3}$,\and
    Jesse J. Lutz$^{3}$,
    Raj Patel$^{4}$,
    Matthew Markham$^{4}$,\and
    Jacob Henshaw$^{3}$, and
    Andy M. Mounce$^{1}$$^{\ast}$.\and
	\small$^{1}$Center for Integrated Nanotechnologies, 1515 Eubank Ave SE, Albuquerque, NM 87123.\and
    \small$^{2}$ Department of Computer Engineering, University of New Mexico \and
	\small$^{3}$Sandia National Laboratory, 1515 Eubank Ave SE, Albuquerque, NM 87123. \and
    \small$^{4}$ Element 6, Fermi Avenue, Harwell OX11 0QRUK. \and
    \small$^{\ast}$ Corresponding author: amounce@sandia.gov
}

\date{}

\begin{document} 


\baselineskip24pt


\maketitle 

\begin{sciabstract}
Solid-state quantum defects (point crystal imperfections with optically addressable spins) are among the most promising platforms for quantum information science due to their long-lived electron and nuclear spin lifetimes. The negatively-charged nitrogen-vacancy (NV$^-$) and negatively-charged group IV vacancies (G4V's) in diamond are some the most promising quantum defects but, the NV$^-$ is highly susceptible to charge noise and has low emission from its zero-phonon line while the G4V's require low operating temperatures. Here we report the physical properties of the newly discovered negatively-charged gallium-vacancy in diamond, GaV$^-$. We find that the GaV$^-$, created through FIB implantation and high-pressure high-temperature annealing, exhibits a combination of the best properties of both the NV$^-$ and G4V's: optically-detectable spin coherence at room temperature ($T_2(293K) = 1.06 \mu$s), a large Debye-Waller factor $\sim$0.4, a short radiative lifetime ($\sim$3.83 ns), strong temperature dependence of the zero-field splitting (16x that of the NV$^-$), and charge-noise-protected D$_{3d}$ symmetry. These features (along with improved performance at low temperature) position the newly discovered GaV$^-$ as a key resource for quantum sensing, networking, and computation. 
\end{sciabstract}

\section{Introduction}

Solid-state quantum defects are crystal substitutions, interstitial atoms, vacancies, or combinations of these that have ideal properties for quantum information science. They are among the top candidates to be quantum resources as part of quantum networks \cite{aharonovich_solid-state_2016, gao_coherent_2015,Awschalom2018,Gordon_2013} because they offer an atomic-footprint solid-state all-in-one solution for quantum memories (nuclear spins), processing (electron spins), and communication (on-demand photons) \cite{Awschalom2018,Gordon_2013,Wolfowicz_2021}.

The negatively charged nitrogen-vacancy, NV$^-$, in diamond has long been the exemplar quantum defect for quantum communication, computation, and sensing \cite{hensen_loophole-free_2015, hermans_qubit_2022, pompili_realization_nodate,Pezzagna_NV_quantum_computer,maze_nanoscale_2008,doherty_nitrogen-vacancy_2013,rondin_magnetometry_2014,abe_tutorial_2018,levine_principles_2019,barry_sensitivity_2020,xu_recent_2023}. However, for quantum communication, the NV$^-$ has a low Debye-Waller factor of $\sim$0.04, leading to a low rate of photon emission from its zero phonon line (ZPL), and a crystal symmetry $C_{3v}$ sensitive to charge noise and spectral diffusion. \cite{aharonovich_diamond-based_2011} These deficiencies led to the prediction and discovery of the group IV-vacancies (G4V's) in diamond, \cite{bradac_quantum_2019} (SiV$^-$ \cite{neu_low-temperature_2013,neu_single_2011,muller_optical_2014}, GeV$^-$ \cite{iwasaki_germanium-vacancy_2015}, SnV$^-$ \cite{trusheim_transform-limited_2020,iwasaki_tin-vacancy_2017}, and PbV$^-$ \cite{wang_low-temperature_2021,trusheim_lead-related_2019,ditalia_tchernij_single-photon_2018,wang_transform-limited_2024}) which have higher Debye-Waller factors and are less susceptible to charge noise due to the split-vacancy crystallographic symmetry ($D_{3d}$). However, the G4V's have a non-ideal electronic and spin structure, with a pair of split ground and excited states arising from the dynamic Jahn-Teller (JT) interaction and spin-orbit coupling, leading to thermally induced spin-decoherence and poor microwave control. \cite{bradac_quantum_2019,qiu_origin_2024}

The SiV$^0$ has recently attracted attention as an alternative to the negatively charged $S=1/2$ G4V's: its 
$S=1$ ground state eliminates the JT distortion leading to long spin coherence at higher temperature than the SiV$^-$ .\cite{thiering_eg_2019} Stabilizing the neutral charge state of the G4V's has proven to be challenging as the neutral charge states are not stable in high-purity diamonds and, at least in the case of SiV$^0$, resonant excitation from the valence band readily converts SiV$^0$ to SiV$^-$ \cite{garciaarellano_photoinduced_2024}. To date, only the SiV$^0$ has been observed, with initial efforts requiring Fermi-level engineering through boron doping, which can degrade the high-purity diamond environment. \cite{rose_observation_2018,green_electronic_2019,zhang_neutral_2023} More recently, it has been shown that surface termination control can reliably convert SiV$^-$ to SiV$^0$ without the need for boron doping.  \cite{zhang_neutral_2023} However, the charge instability of SiV$^0$ as well as the quenching of its photoluminescence at reduced temperatures \cite{zhang_neutral_2023,dhaenens-johansson_optical_2011} remain potential roadblocks to its utility.

The need for alternative quantum defects has led to calculations of the physical properties of quantum defects in silicon \cite{xiong_computationally_2024,xiong_high-throughput_2023}, 2D-materials \cite{Sajid_2D_High_throughput,thomas_substitutional_2024}, and diamond \cite{harris_group-iii_2020, goss_density_2024,qiu_origin_2024}, most recently employing high throughput DFT calculation frameworks for down selecting quantum defect candidates. The GaV$^-$ and other G3V defects in diamond have thus been predicted to be viable alternatives to the G4V's and NV$^-$.\cite{harris_group-iii_2020,goss_density_2024,xiong_identifying_2025} In particular, the GaV$^-$ and other G3V defects were predicted to be thermodynamically stable in the -1 charge state, have a spin-1 ground state, possess $D_{3d}$ symmetry, with a Debye-Waller factor and radiative lifetime on the order of the negative G4V's \cite{harris_group-iii_2020, goss_density_2024, xiong_identifying_2025}, having the combined desirable spin, optical, and spectral properties of the NV$^-$ and the G4V's.

Here, we report the discovery and characterization of the GaV$^-$ center in diamond. We confirm that its properties are consistent with theoretical predictions which stated: an electron spin-1, $D_{3d}$ symmetry, a short radiative lifetime (3.83 $\pm$ 0.10 ns), and a large Debye-Waller factor ($\sim$0.4) \cite{harris_group-iii_2020,goss_density_2024,xiong_identifying_2025}. The combination of the desirable spin and optical properties  positions GaV$^-$ as a key quantum resource for quantum defect based quantum information science. 

\section{The gallium-vacancy in diamond}

\begin{figure}[h]
\centering
\includegraphics[width=1\textwidth]{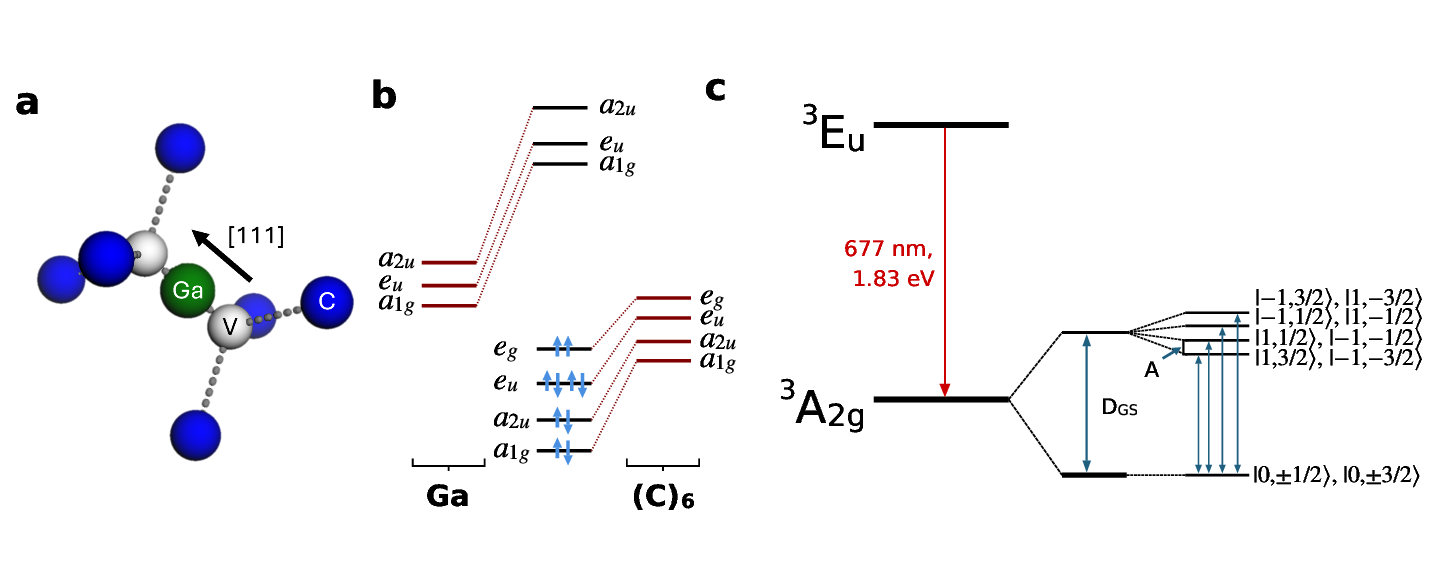}
\caption{(a) Crystal structure of the GaV$^-$ defect showing the split-vacancy configuration with $D_{3d}$ symmetry. (b) Molecular-orbital model for GaV$^-$ illustrating that the relevant orbitals of the GaV$^-$ arise from the dangling bonds of carbon (C)$_6$ which are linear superpositions of the 6 dangling bonds and are labeled by their transformation in the $D_{3d}$ symmetry. The ground state electronic configuration is shown and the first excited state is found by promoting an electron from $e_u$ to $e_g$. (c) Energy level diagram for the many-electron states of the GaV$^-$ arising from the ground state and first excited state electronic configurations with the ground state axial zero field splitting parameter and the isotropic hyperfine splitting parameter are indicated as $D_{GS}$ and $A$, respectively. Energy levels are labeled by the dominant component of the spin eigenstate $\ket{m_s,m_i}$, which are calculated using values from Table \ref{GaV prop}, Equations \ref{Hz} and \ref{Hhf}.}\label{fig1}
\end{figure}

The GaV$^-$ is predicted to have a split-vacancy structure, with $D_{3d}$ symmetry, like the G4V's (Figure \ref{fig1}a). Similar to SiV$^0$ \cite{dhaenens-johansson_optical_2011}, the relevant electron orbitals of the GaV$^-$ defect are expected to arise from the 6 dangling bonds of the neighboring six carbon atoms (Figure \ref{fig1}b). In the $D_{3d}$ symmetry, linear superpositions of these 6 dangling bonds give rise to orbitals that transform as $e_g$, $e_u$, $a_{2u}$ and $a_{1g}$. Likewise, linear superpositions of the 4s and 4p orbitals of gallium give rise to orbitals that transform as $a_{2u}$, $e_u$ and $a_{1g}$. In the presence of the gallium atom, the $(C)_6$ orbitals are lowered in energy and account for the occupied orbitals in the GaV$^-$ defect, while the Ga orbitals are raised in energy and can be neglected. \cite{dhaenens-johansson_optical_2011,harris_group-iii_2020,thiering_eg_2019}

In the hole picture, the ground state occupation is $e^2_g$ and the first excited state occupation is $e_u^1e_g^1$. The $e^2_g$ configuration gives three many-electron states: a triplet $^3A_{2g}$ and two singlet states $^1E_g$ and $^1A_{1g}$. \cite{green_electronic_2019} Based on Hund's rules, observations of the SiV$^0$ \cite{green_electronic_2019,thiel_probing_2019} and our observations of a triplet state in optically detected magnetic resonance (ODMR) (Section \ref{spin}), we assign the triplet $^3A_{2g}$ state to the ground state (Figure \ref{fig1}c). The singlet states, $^1E_g$ and $^1A_{1g}$, are omitted but are higher in energy than $^3A_{2g}$ and below the many-electron states of $e_u^1e_g^1$ \cite{thiel_probing_2019} but could participate in an intersystem crossing which enables ODMR. For the excited state configuration, $e_u^1e_g^1$, which gives rise to 6 many-electron states, $^{1,3}A_{1u}$,  $^{1,3}A_{2u}$ and $^{1,3}E_{u}$, we follow Green, et. al. \cite{green_electronic_2019} and assign $^3E_u$ to the optically active excited state in Figure \ref{fig1}c, which lies 1.83 eV (677 nm, Table \ref{GaV prop}) above $^3A_{2g}$. 


\begin{table}[]
\centering
\begin{tabular}{|c|c|c|}
\hline
Property & Experimental Value & Prediction \\\hline
ZPL (nm) & 677 & 679 \cite{harris_group-iii_2020}, 652 \cite{goss_density_2024} \\
DWF & 0.397 $\pm$ 0.01 & 0.43 \cite{harris_group-iii_2020} \\
$\tau_{rad}$ (ns) & 3.83 $\pm$ 0.10 & - \\
$D$ (MHz) & 3191.0 $\pm$ 0.32 & 3046 \\
$E$ (MHz)  & -3.51 $\pm$ 3.42 & -0.09 \\
$A_{iso}$ (MHz) & -273.2 $\pm$ 0.49 & -223 \cite{goss_density_2024} \\
$A_{ax}$ (MHz) & 6.93 $\pm$ 0.29 & 5 \cite{goss_density_2024}\\\hline
\end{tabular}

\caption{Experimentally measured and calculated optical and spin properties of the GaV$^-$.}\label{GaV prop}
\end{table}

The ground state of the GaV$^-$ defect, $^3A_{2g}$, is a spin-1 triplet, with a zero-field splitting (ZFS), described by the Hamiltonian, $H_{ZFS}$. The nuclear spin of both stable isotopes of Ga, $^{69}$Ga and $^{71}$Ga is 3/2, which yields a hyperfine splitting, described by the Hamiltonian, $H_{hf}$ (Section \ref{spin}). These splittings are shown in Figure \ref{fig1}c, and the relevant parameters are reported in Table \ref{GaV prop}.

\section{Optical properties of the GaV$^-$}\label{optical prop}

\begin{figure}[!h]
\centering
\includegraphics[width=1\textwidth]{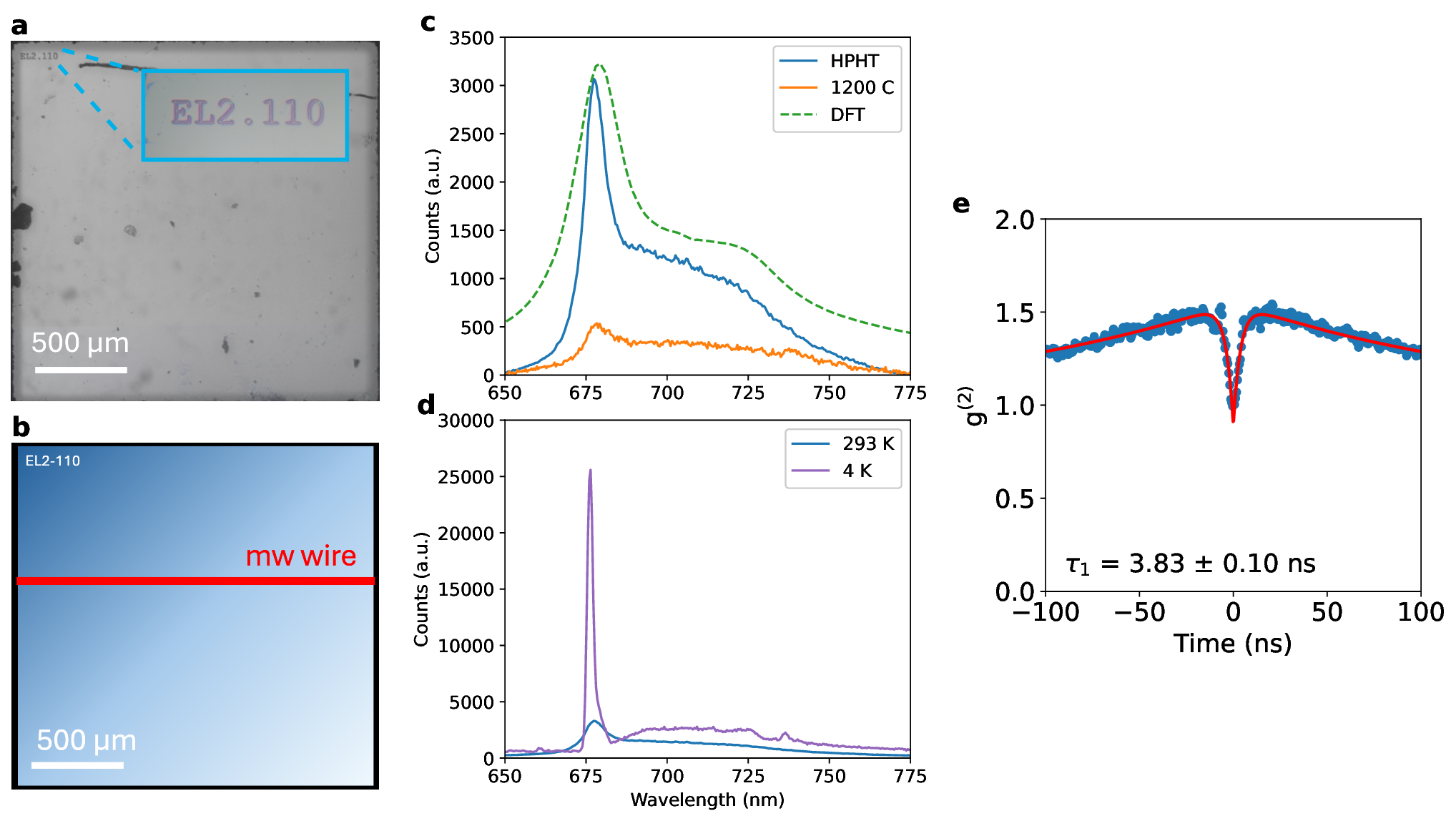}
\caption{(a) White light microscope image of the electronics grade diamond with the FIB mark "EL2-110" with expanded view of the FIB'd mark. (b) Sample schematic showing the location of the FIB marker (EL2-110) and microwave line. The color gradient is reflective of the implant density of Ga from the FIB beam.(c) PL spectra showing GaV spectrum under 532 nm excitation after a vacuum 1200$^{\circ}$C anneal (Orange) \cite{titze_situ_2022}, after HPHT annealing (Blue), and a calculated spectrum from Harris et. al. (Green) \cite{harris_group-iii_2020}. (d) PL spectra showing the GaV at room (293 K) and cryogenic temperature (4 K) under 532 nm excitation. (e) Background corrected second-order photon correlation measurement (blue), g$^{(2)}$ in the middle of the sample showing there are countably few GaV$^-$. We fit it to Equation \ref{g2} and obtain $\tau_1$ = 3.83 $\pm$ 0.10 ns which corresponds to the radiative lifetime of the excited state.}\label{fig2}
\end{figure}

We first discovered the photoluminescence spectrum of the GaV$^-$ by implanting an electronics grade diamond ($<$5 ppb N) from Element 6 with 75 keV gallium ions with a fluence of $3.4 \times 10^{11}$ ions/cm$^2$, and a projected depth of 29.7 nm. After implantation, the diamond was annealed up to 1200$^\circ$C, tri-acid cleaned, and oxygen terminated as described in Section \ref{methods}. Confocal spectroscopy with a 532 nm pump laser revealed a weak photoluminescence peak at 677 nm with a phonon side band extending to approximately 750 nm (Figure \ref{fig2}a). A second sample was unintentionally implanted with $^{69}$Ga during the inscription of a sample identification mark (Figure \ref{fig2}b) using an isotopically enriched 69-gallium focused ion beam at 30 keV. FIB imaging and a beam tail implanted Ga across the diamond sample with a gradient of unknown fluences (Figure \ref{fig2}c), at a projected depth of $\sim$14 nm. 


After implantation, the second diamond sample was annealed at 1200 $^\circ$C following \cite{titze_situ_2022}, then under high temperature and high pressure (HPHT) at 2100 $^\circ$C and 7.7 GPa for 20 minutes (details in Methods). Under 532 nm excitation, we find that the photoluminescence spectrum from GaV$^-$ after HPHT annealing has $\sim$5x greater intensity as compared to a similarly prepared sample annealed at 1200 $^\circ$C (Supplemental Information). The measured value of the ZPL is in close agreement with the prediction of 679 nm given in \cite{harris_group-iii_2020} and of 652 nm given in \cite{goss_density_2024}. Overlaid on the observed HPHT spectrum is the theoretical spectrum taken from \cite{harris_group-iii_2020}, showing the close resemblance between the theoretical and experimental spectra. From the spectrum in Figure \ref{fig2}a, we calculated the Debye-Waller factor (DWF) to be 0.397, which is consistent with the predicted value of $\sim$0.43.\cite{harris_group-iii_2020} Furthermore, at $T = 4 $K, we observe that the ZPL sharpens and the overall brightness of the GaV$^-$ emission increases (Figure \ref{fig2}d), with a linewidth limited by the spectral resolution of our 150 g/mm spectrometer.   

Near the bottom right corner, opposite the FIB marker, the density of implanted Ga atoms is sufficiently low to demonstrate countably few GaV$^-$ defects. In this region, we measure a background corrected second-order correlation, $g^{(2)}$, finding antibunching at $t = 0$ and bunching observed near $t = 0$ (Figure \ref{fig2}e). Thus, we fit g$^{(2)}(t)$ to a three level model using the following equation \cite{aharonovich_photophysics_2010}:

\begin{equation}
    \label{g2}
    g^{(2)}(t) = 1 - \frac{1}{N} + \frac{1-(1+a)e^{-\frac{\lvert t \rvert}{\tau_1}} + a e^{-\frac{\lvert t \rvert}{\tau_2}}}{N},
\end{equation}\label{g2}where N is the number of single-photon emitters, $\tau_1$ is the excited-state lifetime, $\tau_2$ is the shelving state lifetime and a is the ratio of emission from the shelving state to the excited state. Our fit gives  N = 9.5 $\pm$ 0.97, $\tau_1$ = 3.82 $\pm$ 0.1 ns. We assign $\tau_1$, which is on the order of the radiative lifetimes of the G4Vs, to the radiative lifetime of the excited state $^3E_u$.

\section{Spin properties} \label{spin}

\begin{figure}[h]
\centering
\includegraphics[width=1\textwidth]{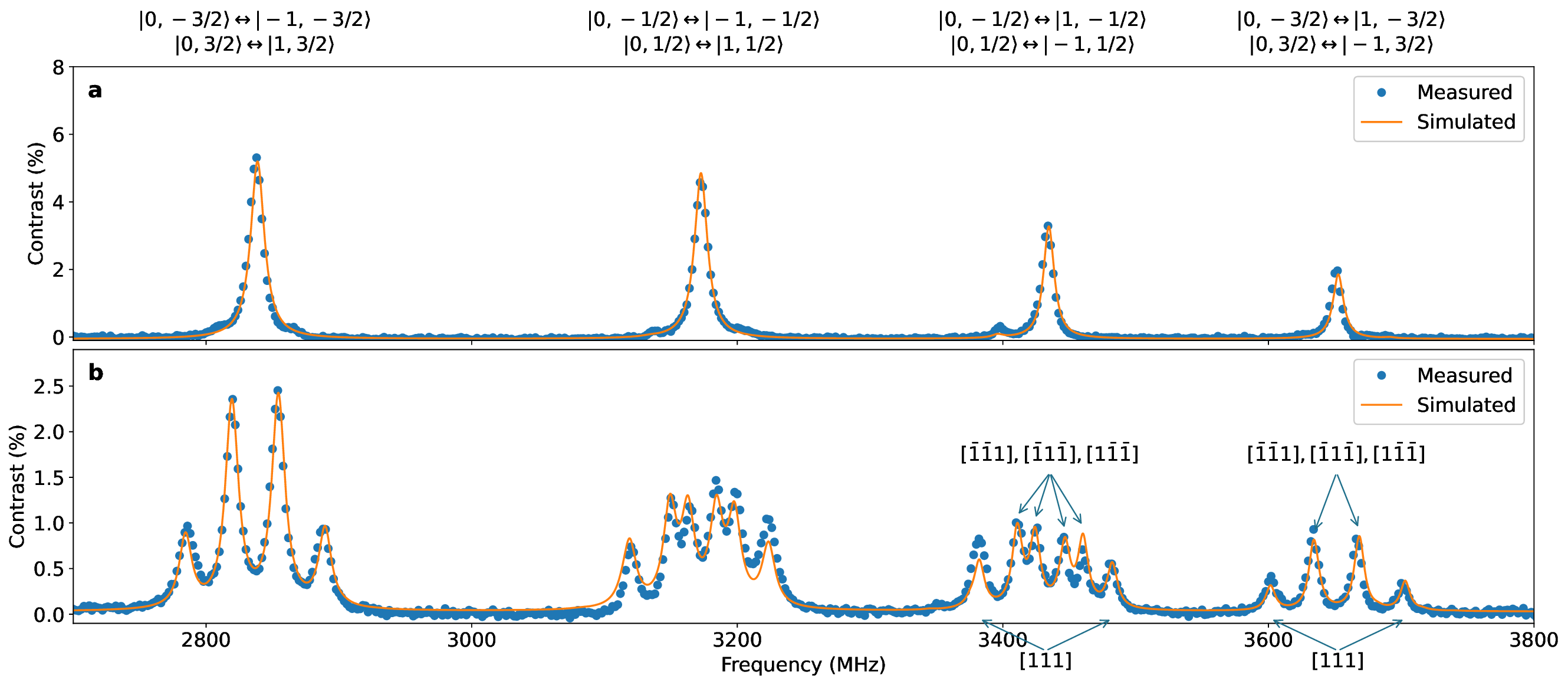}
\caption{(a) Observed (blue) and simulated (orange) zero-field ODMR spectrum at 293 K under 637 nm excitation. (b) In-field ODMR spectrum (blue) with $B_0$ = 19 G and oriented along the $\langle111\rangle$ symmetry axis. Labels indicate the spin transitions and crystallographic orientations\label{fig3}.}
\end{figure}

We find that the GaV$^-$ has an ODMR response at room temperature with  4 peaks in zero magnetic field under 637 nm excitation (Figure \ref{fig3}a) which we label according to the spin energy levels shown in Figure \ref{fig1}c. The presence of 4 peaks is consistent with an $S= 1$ and $I=3/2$ spin system, where the degenerate ZFS electron transitions ($\ket{0} \rightarrow \ket{\pm1}$) are split into $2I + 1 = 4$ transitions due to hyperfine coupling with the nucleus of spin $I=3/2$. 

Like the NV and G4V's, the symmetry of the GaV$^-$ gives rise to four different orientations relative to the crystallographic axes: $[111]$, $[\overline{1}\overline{1}1]$, $[\overline{1}1\overline{1}]$, $[1\overline{1}\overline{1}]$. In zero magnetic field, the spin energy levels of these orientations are all degenerate. However, with the application of a magnetic field along the [111] direction, the [111] will have the largest energy splitting while the other three directions will be degenerate with a smaller splitting. Indeed, with a 19 G magnetic field applied along the $[111]$ orientation of the diamond (see Section \ref{methods}), the 4 peaks measured in the zero-field case split into 20 peaks (Figure \ref{fig3}b).

To fit the observed ODMR spectrum, we consider that the spin energy levels of the GaV$^-$ are primarily governed by three Hamiltonians: the Zeeman ($\mathcal{H}_Z$), zero field splitting ($\mathcal{H}_{zfs}$), and hyperfine ($\mathcal{H}_{hf}$) Hamiltonians. We define the Zeeman Hamiltonian with an isotropic $g$-factor as:

\begin{equation}\label{Hz}
\mathcal{H}_Z(\vec{B_0}) = \frac{g_{iso}\mu_B}{2}\vec{B_0}\cdot\hat{S},
\end{equation} where $g_{iso}$ is the electron g-factor, $\mu_B$ is the Bohr magneton, $\vec{B_0}$ is the external magnetic field vector, $\hat{S}$ is the electron spin operator. We use the zero field splitting Hamiltonian defined as:

\begin{equation}\label{Hzfs}
\mathcal{H}_{zfs}(D, E) = D (S_z^2 - \frac{1}{3}S(S+1)) + E (S_x^2 - S_y^2),
\end{equation} where $D$ and $E$ are the axial and transverse zero-field splitting parameters, S is the total electron spin and $S_{x,y,z}$ are the electron Pauli matrices for $S = 1$. Finally, we consider both isotropic and axial components of the hyperfine Hamiltonian defined as:

\begin{equation}\label{Hhf}
\mathcal{H}_{hf}(A_{iso}, A_{ax}) = 
    (A_{iso} + 2A_{ax})S_zI_z + (A_{iso} - A_{ax})(S_xI_x + S_yI_y),
\end{equation} where $A_{iso}$ is the isotropic hyperfine constant, $A_{ax}$ is the axial hyperfine constant, $I_{x,y,z}$ are the nuclear spin Pauli matrices for $I = 3/2$ and $S_{x,y,z}$ is defined above.

We find the energy eigenvalues for the spin levels governed by the sum of Equations \ref{Hz}-\ref{Hhf} with $B_0$ = 0 and $B_0$ = 19 G along $[111]$, respectively to fit the ZFS parameters, $D$ and $E$, and the hyperfine parameters, $A_{iso}$ and $A_{ax}$, for both the zero field and in-field ODMR spectra. Using the difference between energy eigenvalues and the transition probabilities from Fermi's golden rule, we constructed a simulated ODMR spectrum (Fig. \ref{fig3}, orange) for the four different orientations of a $\langle111\rangle$ aligned defect to use in curve fitting the spectra (see Section \ref{methods}). The parameters extracted from fitting the zero field and in-field spectra are reported in the SI with their averaged values reported in Table \ref{GaV prop}. Further, we find that the experimentally measured values of the hyperfine couplings, $A_{iso}$ and $A_{ax}$, show good agreement with the theoretically predicted values for a GaV$^-$ defect with D$_{3d}$ symmetry. \cite{goss_density_2024}

We also measured the zero-field ODMR at 4 K (see Supplementary Information, Fig. S1). We found that the axial zero-field splitting parameter, D, increased by 83 MHz from 3.197 GHz to 3.28 GHz while E, $A_{iso}$ and $A_{ax}$ remained constant to within the fitting error. We note that this change in D over the temperature range from 4 K to 293 K is nearly 16x larger than the $\sim$5 MHz shift observed for NV$^-$ over the same range.\cite{temp_sens_Gali} A table  of our fit parameters for spectra $T = 4, 295$K and $B_0 = 0, 19$ G are provided in Supplementary Information Table S1.

\begin{figure}[h!]
\centering
\includegraphics[width=1\textwidth]{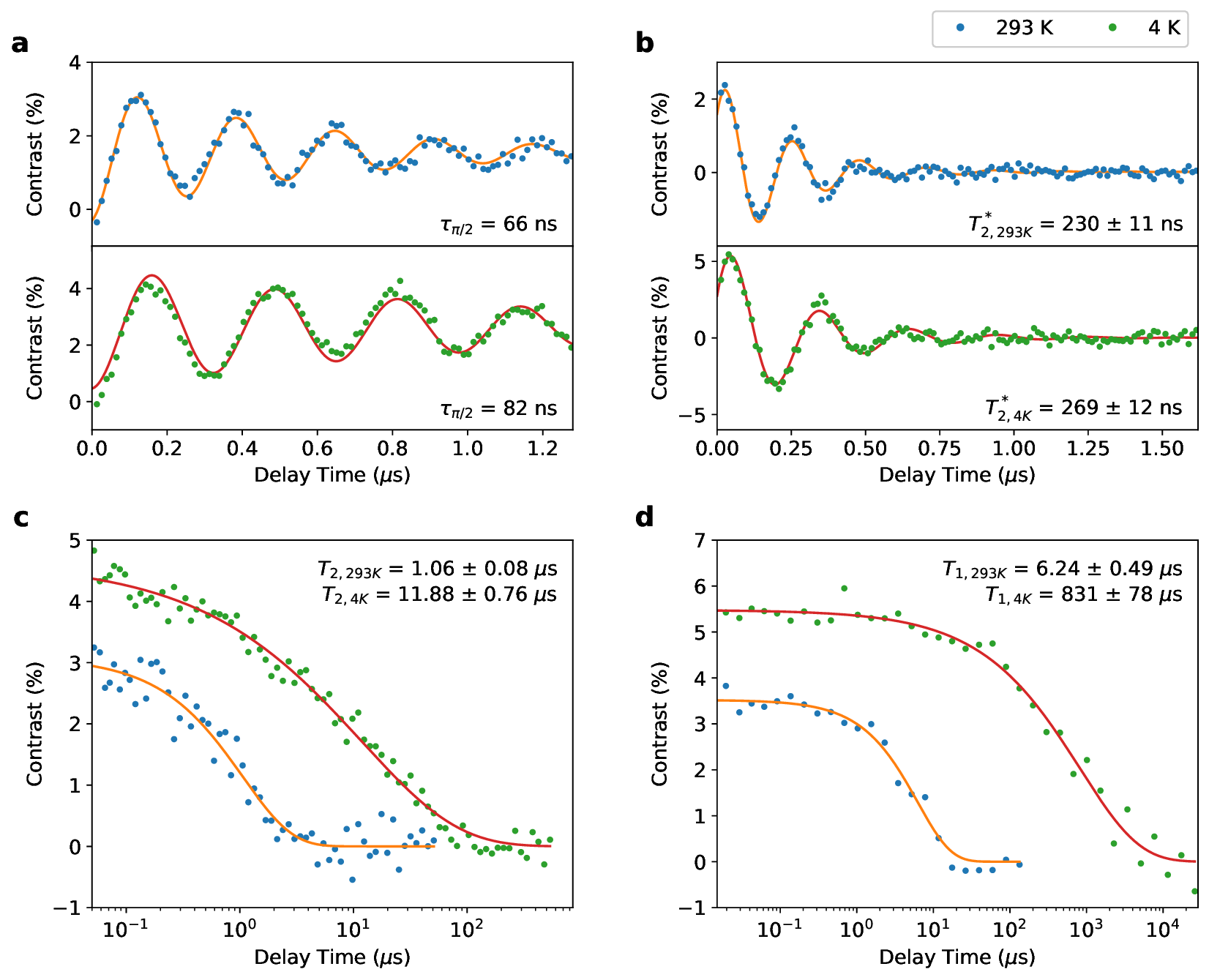}
\caption{Measured (a) Rabi oscillations, (b) Ramsey interferometry, (c) Hahn echo, and (d) inversion recovery at 293 K (blue) and 4 K (green) with fits (293 K, orange, 4 K, red).} \label{fig4}
\end{figure}



We measured the room-temperature (293 K) and 4 K dynamic spin properties ($T_2^*$, $T_2$, and $T_1$) of the GaV$^-$ using the highest intensity ODMR peak at 2837 MHz (2920 MHz at 4 K). We demonstrate coherent control of the GaV$^-$ spins at room temperature and 4 K by measuring Rabi oscillations (Figure \ref{fig4}a) to extract the appropriate $\frac{\pi}{2}$ pulse for the remaining experiments. Using Ramsey interferometry (Figure \ref{fig4}b) we measure the inhomogeneous spin dephasing time, $T_{2,293K}^*$, to be 230 $\pm$ 11 ns and $T_{2,4K}^*$, to be 269 $\pm$ 12 ns. To measure the spin dephasing time, $T_2$, we used a Hahn echo sequence (Figure \ref{fig4}c) and find that $T_{2,293K}$ = 1.06 $\pm$ 0.08 $\mu$s and $T_{2,4K}$ = 11.88 $\pm$ 0.76 $\mu$s. At low temperatures, the Hahn echo decay signal, $S(t)$, is found to follow a stretched exponential, $S(t) = A \cdot exp[-(t/T_2)^n]$. We extract a stretching factor of n $\cong$ 1/2, which indicates that the spatial configuration of the spins in the spin bath changes over time, an effect common for near-surface NV$^-$ \cite{doi:10.1073/pnas.2422542122}. Given the proximity of GaV$^-$ to the surface, a similar effect to what is observed in NV$^-$ is likely occurring. Finally, to measure $T_1$ we perform an inversion recovery measurement (Figure \ref{fig4}d) and find that $T_{1,293K}$ = 6.99 $\pm$ 0.31 $\mu$s and $T_{1,4K}$ = 814 $\pm$ 80 $\mu$s. Once again, at low temperatures, the signal decay follows a stretched exponential with n $\cong$ 1/2. This suggests that the decoherence mechanism for $T_1$ and $T_2$ at 4 K is likely the same \cite{rose_observation_2018}. We note that the proximity to the surface and the observed decay behavior suggest that the measured coherence lifetimes are on the lower end of the distribution of possible lifetimes for GaV$^-$. Further details on measurement techniques are provided in the Methods section.

\section{Discussion}

Our discovery of the GaV$^-$ presents exciting possibilities for quantum information science. For quantum networking and communication, the GaV$^-$ has the best features of both NV's and the G4V's. Like the NV$^-$, the GaV$^-$ has $S=1$ resulting in spin coherence at high temperatures. Similar to G4V's, the GaV$^-$ is protected from charge noise and has a large Debye-Waller factor.  Further, as opposed to the SiV$^0$, the GaV$^-$ is charge stable when implanted in intrinsic diamond, not requiring charge engineering. Additional investigation into the charge dynamics of the GaV$^-$ and the optimal formation conditions could present the GaV$^-$ as a new front-runner as a quantum resource for quantum communication. 

As a quantum sensor, the GaV$^-$ is more promising than many other defects in diamond, but the measured spin coherence times are shorter than the NV$^-$. In the present work, the measured T$_2^*$ is at least an order of magnitude shorter than the T$_2^*$ achievable in NV$^-$ centers, which limits its sensitivity as a static magnetic field sensor. However, due to the uncontrolled fluence and shallow formation it's unclear what the limits of $T_2^*$ will be until further investigation is performed. \cite{NV_sensitivity_Barry} On the other hand, the performance of the NV$^-$ as a single-spin sensor can degrade at low temperature or in vacuum due to changes in charge state with complicated dynamics between the zero- and negative- charge state.\cite{kumar2026stabilisation}   Alternatively, our measurements of the GaV$^-$ only show one stable charge state, which may prove more robust to harsh environments. The prospects for GaV$^-$ for temperature sensing are also quite intriguing. The temperature sensitivity, $\eta$, is inversely proportional to $\frac{dD}{dT}$ \cite{temp_sens_Awschalom} and we find that  $\Delta D = D(T=293K) - D(T=4K) \sim$83 MHz, relative to the NV's $\Delta D_{NV} \sim $5 MHz, which indicates a potential for 16x better temperature sensitivity. \cite{temp_sens_Gali} 

\section{Conclusions}

In this work, we report the discovery of the GaV$^-$ and that it has optically detectable spin coherence at room temperature, a short radiative lifetime of 3.83 $\pm$ 0.10 ns, a large Debye-Waller factor of 0.397, and $D_{3d}$ symmetry. Overall, the GaV$^-$ defect combines the desirable properties of the NV$^-$ and G4V's, likely overcoming the deficiencies of these quantum defects. While additional investigation of the GaV$^-$ is necessary and ongoing, the results reported in the current work position it as a key quantum resource for defect-based quantum information science applications, with strong promise for use as a quantum resource in a quantum network or as a quantum sensor. 

\section{Methods}\label{methods}

\subsection{GaV Formation}

Initial GaV$^-$ optical spectra were collected from an electronics grade diamond from Element 6, implanted with 75 keV Ga ions at a fluence of $3.4\times10^{11}$ ions/cm$^2$. For our FIB-implanted diamond, used in all spin measurements, we used an electronic grade diamond ($<$5 ppb N) from Element 6 and inscribed a serial number marker "EL2-110" in a FEI Nova Nanolab 500 Dualbeam System using a 30 keV isotopically purified 69-Ga using a Gallium-69 LMIS source from Applied Beams. 

Both of these samples were cleaned in a piranha solution (3:1 H$_2$SO$_4$:H$_2$O$_2$) to remove any organic residue. The sample was then annealed at $1\times10^{-8}$ Torr with the following temperature progression. The sample was ramped to 400 $^{\circ}$C, held for 4 hours, ramped to 800 $^{\circ}$C, held for 4 hours, ramped to 1200 $^{\circ}$C and held for 2 hours, then cooled to room temperature. The ramp rate for heating and cooling was a constant 100 $^{\circ}$C/h. After the anneal, the sample was cleaned in a triacid bath of 1:1:1 H$_2$SO$_4$:HNO$_3$:HClO$_4$ in a boiling flask at 250 $^{\circ}$C that was refluxed with a water-cooled condenser for 1.5 hours. Following the triacid cleaning, the surface was oxygen-terminated with a UV ozone cleaner (Uvocs T10X10) for 100 minutes. 

Prior to the high-pressure and high-temperature anneal, the FIB-implanted sample was cleaned in a piranha solution and then HPHT annealed at 2100 $^{\circ}$C and 7.7 GPa for 20 minutes by Element 6. After the anneal, the sample was again cleaned in a triacid bath and the surface was oxygen terminated. Another FIB-implanted reference sample was similarly inscribed with FIB before being annealed at 1200 $^\circ$C, cleaned and oxygen terminated as described above.

\subsection{Optical Measurements}

The room temperature and 4 K ODMR measurements were performed in a Montana cryostat using a home-built microscope similar to the one described in Henshaw et. al \cite{henshaw_nanoscale_2022}. The 637 nm excitation for ODMR spectra, Rabi oscillations, Ramsey interferometry, and Hahn echo measurements was provided by a 50 mW Thorlabs SM Fiber-pigtailed laser diode pulsed using a Gooch $\&$ Housego acousto-optic modulator (AOM) which provides 20 mW of laser power to the microscope. For the $T_1$ measurement, we used a Cobolt 06-01 Series 638 nm laser from Hubner Photonics to eliminate light leakage from the AOM. 

The AOM-coupled laser was focused near the back focal plane of the 100x 0.8 NA objective with a 400 mm lens illuminating a 3 $\mu$m diameter spot on the diamond. While the Cobolt 06-01 laser was not focused on the back aperture to expand it on the sample surface, the power density at the sample surface was matched to the AOM-coupled setup, while using a confocally limited spot. We collect and reflect the light on a 505 nm short pass dichroic (505DMSP), 650 nm long pass dichroic (650 DMLP) and two 650 nm long-pass filters to eliminate laser leakage. This light is focused by a 50 mm lens through a 100 micron pinhole (30-micron pinhole while using the Cobolt 06-01 laser) to eliminate unwanted background before being recollimated by a 50 mm lens. The collected light is focused into a 105-micron diameter fiber that directs the light to an Excelitas Technologies single photon counting module (SPCM). The pulses from the SPCM were counted using an RFSoC 4x2 and the Quantum instrumentation control kit – Defect arbitrary waveform generator (QICK-DAWG). \cite{riendeau_quantum_2023}  For the pulsed measurements, we define contrast as the difference between the -z projection ($\ket{\pm1}$) and the +z projection ($\ket{0}$) on the Bloch sphere. This contrast is converted to $\%$ contrast by normalizing to a reference that has been initialized to the $\ket{0}$ state by laser pumping. In contrast to NV$^-$, based on Figure \ref{fig3}, the $\ket{\pm1}$ state is brighter under laser pumping than the $\ket{0}$ is, yielding a positive contrast.

The magnetic field for the in-field measurements presented in the text was applied using a Ferronato BH300-3-A Helmholtz coil. The magnetic field direction was tuned to the [111] direction using a sample with a thin surface layer of NV$^-$. The magnetic field strength was determined using the manufacturer's specifications for the Helmholtz coil.

Optically detected magnetic resonance was performed using the RFSoC 4x2 and QICK-DAWG to generate the required microwave pulses and control the AOM/Hubner Photonics laser to generate the required laser pulses for Ramsey interferometry, Hahn echo and inversion recovery experiments. The microwave pulses are amplified by a Mini-circuits ZHL-16W-43+ and delivered to the sample through a wire (Figure \ref{fig2}c) on top of the sample.

The room temperature second-order photon correlation ($g^{(2)}$) measurements were performed in a home-built confocal microscope described in \cite{titze_situ_2022}. A fiber-coupled 532 nm continuous wave laser diode at 12 mW (Thorlabs) was reflected using a 532 nm dichroic (Di02-R532-25x36) into an objective. The collected photoluminescence (PL) was filtered back through the 532 nm dichroic, focused by a 180 mm lens through a 50 micron pinhole, recollimated and filtered through a 550 nm LP filter and 650 nm LP filter. The collected PL was directed into a free-space Hanbury-Brown-Twiss (HBT) interferometer.

The 4 K and room temperature optical spectra were collected in a CryoAdvance from Montana instruments equipped with a Montana Cryo-optic and xyz piezo stack from Attocube. The 532 nm excitation was provided by a 5 W Lighthouse Photonics Sprout laser which is shared across multiple setups. Although we find that 637 nm excitation is $\sim$7x more efficient at exciting the GaV$^-$, for clarity, we used 532 nm to collect the spectra because under 637 nm excitation the 1st order diamond Raman line is at 696 nm. We also note that an ODMR response is evident under 532 nm and 637 nm excitation. A Gooch $\&$ Housego acousto-optic modulator (AOM) which provides 150 mW of laser power to the setup directed into the cryostat with a 505 nm dichroic (505 DMSP) where it focuses on the sample with a typical spot size of $\sim$3 microns. The light collected by the cryo-optic is filtered through a 550 nm dichroic (550 DMLP), 533 nm notch filter (533-14 NF) and 550 nm long-pass filter to eliminate the laser line. The light is then focused through a 100-micron pinhole and recollimated by two 50 mm lenses to reduce collected emission from the underlying board. The light was then coupled into 105 micron fiber directing light into a Princeton Instruments HRS-300 spectrometer, equipped with a PiXiS-100 and a WITec alpha300 series Raman microscope. Custom printed circuit boards (PCB) for these experiments were manufactured using an LPKF Protolaser U4.

\subsection{Electronic structure simulations}

Utilizing Quantum Espresso (QE) V7.5 \cite{Giannozzi2009,Giannozzi2017}, the $S=1$ GaV$^{\textendash{}}$ defect center was introduced to a 64-atom diamond supercell within a 7.136032 \AA\ unit cell. Relaxation of atomic positions was performed using spin-polarized density functional theory with the PBE functional \cite{Perdew1996}, V1.0 PBE optimized norm-conserving Vanderbilt PBE pseudopotentials \cite{Hamann2013}, and a wavefunction energy cuttoff of 60 eV. Brillouin-zone sampling was performed at the Gamma point and symmetry was disabled. A subsequent converged HSE06 wavefunction \cite{heyd2003JCP,heyd2006JCP} was post-processed using pyZFS \cite{ma2020} to generate a spin-spin contribution to the ZFS axial asymmetry of $D^{SS}=1354$ MHz. The spin-spin D tensor was validated using VASP V6.6.0 \cite{kresse1993PRB,kresse1994PRB,kresse1994JPCM,kresse1996CMS,kresse1996PRB} within the PAW formalism \cite{blochl1994PRB,kresse1999PRB}, where GaV- was introduced to a 215-atom diamond supercell within a 10.646060 \AA\ unit cell. These calculations included d-orbitals as valence with a plane-wave cutoff of 520 eV. The ZFS D tensor was generated \cite{marsman2010JCP} to obtain $D^{SS} = 1280$ MHz, in reasonable agreement with the pyZFS value. 

The spin-orbit contribution to the D tensor was computed using Orca V6.1 \cite{ORCA6,Neese2022}. To generate the 54C+1Ga+60H cluster, the relaxed QE supercell was trimmed to generate a C2h-symmetric cluster, it was hydrogen terminated, and only the H atoms were relaxed at the PBE/def2-SVD level. In addition to the auxiliary RI correlated basis set cc-pVDZ/C, we deployed cc-pVTZ-DK (cc-pVDZ-DK) for Ga and its 6 nearest-neighbor C atoms (all other C and H atoms). A complete active space strongly-contracted N-electron valence 2nd-order perturbation theory (CAS-SC-NEVPT2) calculation \cite{Kollmar2021,HelmichParis2021,Guo2021,Lang2019,Kollmar2019} was then performed utilizing the Resolution-of-the-Identity (RI) approximation \cite{Neese2003}. We utilized an active space of 8 electrons in 5 orbitals, including highest-occupied molecular orbitals having significant hybridization with the Ga 4s, 4p, and 3d atomic orbitals. Relativity was treated approximately through the 2nd-order Douglas-Kroll-Hess transformation and the spin-orbit mean-field approximation. The ZFS was computed to be $D^{SO} = 1692$ MHz using quasi-degenerate perturbation theory with NEVPT2 diagonal energies \cite{Ganyushin2006} including the ground-state triplet and a manifold of the three lowest-lying singlet states. Using consistent underlying structural models of the CAS-SC-NEVPT2 spin-orbit contribution and the pyZFS spin-spin contribution, we found a total ZFS value of $D^{\mathrm{tot}}=D^{SO}+D^{SS} = 3046$ MHz.

\subsection{Spectrum Simulation}
We construct the total Hamiltonian for both the four possible orientations of the GaV$^-$ in zero-field and applied magnetic field using equations \ref{Hz}, \ref{Hzfs} and \ref{Hhf} and find the energy eigenvalues and eigenvectors using the Quantum Toolbox in Python (QuTiP) and numpy. We then found the spin transitions by calculating the energy differences between energy eigenvalues $\Delta E_{i, j}$ = $|E_i - E_j|$ along with the transition probabilities using Fermi's golden rule:

\begin{equation}
    \mathcal{P}_{i, j}(e_i, e_j) \propto |\bra{e_j} S_\pm \ket{e_i}|^2
\end{equation}

where $e_{i, j}$ are the eigenvectors corresponding to $E_{i, j}$. We simulate our ODMR spectra assuming a Lorentzian line shape with $\Delta E_{i, j}$ as the position of transitions with $\mathcal{P}_{i,j}$ $\neq$ 0, a fit amplitude, and a fit linewidth. The fit amplitude allows us to account for the observed deviation in the peak amplitudes from $\mathcal{P}_{i,j}$, which predicts the same amplitude for the outer set of peaks and the inner set of peaks, respectively. In the zero-field case, we allow the zero-field parameters, D and E and the hyperfine parameters A$_{iso}$ and A$_{ax}$ to vary and perform a least-squares regression to find the optimal values. For the in-field case, we include the magnetic field vector B and allow it to vary as well and repeat the least-squares regression to find the optimal values of A$_{iso}$, A$_{ax}$, D, E, and $\vec{B}_0$. For the magnetic field strength and direction, our fit gives $\lvert \vec{B_0} \rvert$ = 19 $\pm$ 0.009 G with a direction differing from the [111] direction by $<$0.06$^{\circ}$, which is consistent with the calculated strength, 19 G, and direction, [111].



\section{Acknowledgments}

We acknowledge S. Sun and K. Ngan for early discussions and V. Acosta for discussions related to the electronic structure. This work was funded, in part, by the Laboratory Directed Research and Development Program and performed, in part, at the Center for Integrated Nanotechnologies, an Office of Science User Facility operated for the U.S. Department of Energy (DOE) Office of Science. NSF supported T. Dudley  through the Quantum Leap Challenge Institutes (Quantum Systems through Entangled Science and Engineering) under Grant OMA-2016244. Sandia National Laboratories is a multimission laboratory managed and operated by National Technology \& Engineering Solutions of Sandia, LLC, a wholly owned subsidiary of Honeywell International, Inc., for the U.S. DOE’s National Nuclear Security Administration under contract DE-NA-0003525. The views expressed in the article do not necessarily represent the views of the U.S. DOE or the United States Government.

\bibliography{bibliography}

@misc{kumar2026stabilisation,
      title={Stabilisation of NV centres in diamond nanopillars at low temperature}, 
      author={Ravi Kumar and Saksham Mahajan and Felix Donaldson and Leonardo Santoni and Aysha A. Riaz and Gediminas Seniutinas and Felipe Favaro de Oliveira and Anna Regoutz and Fabrizia Foglia and Siddharth Dhomkar and John J. L. Morton},
      year={2026},
      eprint={2604.27826},
      archivePrefix={arXiv},
      primaryClass={physics.app-ph},
      url={https://arxiv.org/abs/2604.27826}, 
}

@article{gao_coherent_2015,
	title = {Coherent manipulation, measurement and entanglement of individual solid-state spins using optical fields},
	volume = {9},
	number = {6},
	journal = {Nat. Photon.},
	author = {Gao, W. B. and Imamoglu, A. and Bernien, H. and Hanson, R.},
	year = {2015},
	pages = {363},
}

@article{trusheim_lead-related_2019,
	title = {Lead-related quantum emitters in diamond},
	volume = {99},
	number = {7},
	journal = {Phys. Rev. B},
	author = {Trusheim, Matthew E. and Wan, Noel H. and Chen, Kevin C. and Ciccarino, Christopher J. and Flick, Johannes and Sundararaman, Ravishankar and Malladi, Girish and Bersin, Eric and Walsh, Michael and Lienhard, Benjamin and Bakhru, Hassaram and Narang, Prineha and Englund, Dirk},
	year = {2019},
	pages = {075430}
}

@article{aharonovich_diamond-based_2011,
	title = {Diamond-based single-photon emitters},
    journal = {Rep. Prog. Phys.},
	author = {Aharonovich, I and Castelletto, S and Simpson, D A and Su, C-H and Greentree, A D and Prawer, S},
	volume = {74},
	issn = {0034-4885, 1361-6633},
	url = {https://iopscience.iop.org/article/10.1088/0034-4885/74/7/076501},
	doi = {10.1088/0034-4885/74/7/076501},
	number = {7},
	urldate = {2025-03-13},
	month = jul,
	year = {2011},
	pages = {076501},
}

@article{ditalia_tchernij_single-photon_2018,
	title = {Single-{Photon} {Emitters} in {Lead}-{Implanted} {Single}-{Crystal} {Diamond}},
	volume = {5},
	copyright = {http://pubs.acs.org/page/policy/authorchoice\_termsofuse.html},
	issn = {2330-4022, 2330-4022},
	url = {https://pubs.acs.org/doi/10.1021/acsphotonics.8b01013},
	doi = {10.1021/acsphotonics.8b01013},
	language = {en},
	number = {12},
	urldate = {2025-03-13},
	journal = {ACS Photonics},
	author = {Ditalia Tchernij, S. and Lühmann, T. and Herzig, T. and Küpper, J. and Damin, A. and Santonocito, S. and Signorile, M. and Traina, P. and Moreva, E. and Celegato, F. and Pezzagna, S. and Degiovanni, I. P. and Olivero, P. and Jakšić, M. and Meijer, J. and Genovese, P. M. and Forneris, J.},
	month = dec,
	year = {2018},
	pages = {4864--4871},
}

@article{neu_low-temperature_2013,
	title = {Low-temperature investigations of single silicon vacancy colour centres in diamond},
	volume = {15},
	number = {4},
	journal = {New J. Phys.},
	author = {Neu, Elke and Hepp, Christian and Hauschild, Michael and Gsell, Stefan and Fischer, Martin and Sternschulte, Hadwig and Steinmüller-Nethl, Doris and Schreck, Matthias and Becher, Christoph},
	year = {2013},
	pages = {043005}
}

@article{bradac_quantum_2019,
	title = {Quantum nanophotonics with group {IV} defects in diamond},
	volume = {10},
	number = {1},
	journal = {Nat. Commun.},
	author = {Bradac, Carlo and Gao, Weibo and Forneris, Jacopo and Trusheim, Matthew E. and Aharonovich, Igor},
	year = {2019},
	pages = {5625}
}

@article{aharonovich_solid-state_2016,
	title = {Solid-state single-photon emitters},
	volume = {10},
	doi = {10.1038/nphoton.2016.186},
	number = {10},
	journal = {Nat. Photon.},
	author = {Aharonovich, Igor and Englund, Dirk and Toth, Milos},
	year = {2016},
	pages = {631},
}

@article{titze_situ_2022,
	title = {In {Situ} {Ion} {Counting} for {Improved} {Implanted} {Ion} {Error} {Rate} and {Silicon} {Vacancy} {Yield} {Uncertainty}},
	volume = {22},
	copyright = {https://doi.org/10.15223/policy-029},
	issn = {1530-6984, 1530-6992},
	url = {https://pubs.acs.org/doi/10.1021/acs.nanolett.1c04646},
	doi = {10.1021/acs.nanolett.1c04646},
	language = {en},
	number = {8},
	urldate = {2025-03-13},
	journal = {Nano Letters},
	author = {Titze, Michael and Byeon, Heejun and Flores, Anthony and Henshaw, Jacob and Harris, C. Thomas and Mounce, Andrew M. and Bielejec, Edward S.},
	month = apr,
	year = {2022},
	pages = {3212--3218},
}

@article{Ma2020,
  title = {PyZFS: A Python package for first-principles calculations of zero-field splitting tensors},
  volume = {5},
  ISSN = {2475-9066},
  url = {http://dx.doi.org/10.21105/joss.02160},
  DOI = {10.21105/joss.02160},
  number = {47},
  journal = {Journal of Open Source Software},
  publisher = {The Open Journal},
  author = {Ma,  He and Govoni,  Marco and Galli,  Giulia},
  year = {2020},
  month = Mar,
  pages = {2160}
}

@article{Perdew1996,
  title = {Generalized Gradient Approximation Made Simple},
  volume = {77},
  ISSN = {1079-7114},
  url = {http://dx.doi.org/10.1103/PhysRevLett.77.3865},
  DOI = {10.1103/physrevlett.77.3865},
  number = {18},
  journal = {Physical Review Letters},
  publisher = {American Physical Society (APS)},
  author = {Perdew,  John P. and Burke,  Kieron and Ernzerhof,  Matthias},
  year = {1996},
  month = Oct,
  pages = {3865–3868}
}

@article{Hamann2013,
  title = {Optimized norm-conserving Vanderbilt pseudopotentials},
  volume = {88},
  ISSN = {1550-235X},
  url = {http://dx.doi.org/10.1103/PhysRevB.88.085117},
  DOI = {10.1103/physrevb.88.085117},
  number = {8},
  journal = {Physical Review B},
  publisher = {American Physical Society (APS)},
  author = {Hamann,  D. R.},
  year = {2013},
  month = Aug 
}

@article{ORCA6,
author = {Neese, Frank},
title = {Software Update: The ORCA Program System—Version 6.0},
journal = {Wiley Interdiscip. Rev. Comput. Mol. Sci.},
volume = {15},
number = {2},
pages = {e70019},
doi = {10.1002/wcms.70019},
year = {2025}
}

@article{Neese2003,
  title = {An improvement of the resolution of the identity approximation for the formation of the Coulomb matrix},
  volume = {24},
  ISSN = {1096-987X},
  url = {http://dx.doi.org/10.1002/jcc.10318},
  DOI = {10.1002/jcc.10318},
  number = {14},
  journal = {Journal of Computational Chemistry},
  publisher = {Wiley},
  author = {Neese,  Frank},
  year = {2003},
  month = Aug,
  pages = {1740–1747}
}

@article{Ganyushin2006,
  title = {First-principles calculations of zero-field splitting parameters},
  volume = {125},
  ISSN = {1089-7690},
  url = {http://dx.doi.org/10.1063/1.2213976},
  DOI = {10.1063/1.2213976},
  number = {2},
  journal = {The Journal of Chemical Physics},
  publisher = {AIP Publishing},
  author = {Ganyushin,  Dmitry and Neese,  Frank},
  year = {2006},
}

@article{Kollmar2019,
  title = {A perturbation‐based super‐CI approach for the orbital optimization of a CASSCF wave function},
  volume = {40},
  ISSN = {1096-987X},
  url = {http://dx.doi.org/10.1002/jcc.25801},
  DOI = {10.1002/jcc.25801},
  number = {14},
  journal = {Journal of Computational Chemistry},
  publisher = {Wiley},
  author = {Kollmar,  Christian and Sivalingam,  Kantharuban and Helmich‐Paris,  Benjamin and Angeli,  Celestino and Neese,  Frank},
  year = {2019},
  month = Feb,
  pages = {1463–1470}
}

@article{Lang2019,
  title = {Spin-dependent properties in the framework of the dynamic correlation dressed complete active space method},
  volume = {150},
  ISSN = {1089-7690},
  url = {http://dx.doi.org/10.1063/1.5085203},
  DOI = {10.1063/1.5085203},
  number = {10},
  journal = {The Journal of Chemical Physics},
  publisher = {AIP Publishing},
  author = {Lang,  Lucas and Neese,  Frank},
  year = {2019},
  month = Mar 
}

@article{Guo2021,
  title = {Approximations of density matrices in N-electron valence state second-order perturbation theory (NEVPT2). I. Revisiting the NEVPT2 construction},
  volume = {154},
  ISSN = {1089-7690},
  url = {http://dx.doi.org/10.1063/5.0051211},
  DOI = {10.1063/5.0051211},
  number = {21},
  journal = {The Journal of Chemical Physics},
  publisher = {AIP Publishing},
  author = {Guo,  Yang and Sivalingam,  Kantharuban and Neese,  Frank},
  year = {2021},
}

@article{HelmichParis2021,
  title = {An improved chain of spheres for exchange algorithm},
  volume = {155},
  ISSN = {1089-7690},
  url = {http://dx.doi.org/10.1063/5.0058766},
  DOI = {10.1063/5.0058766},
  number = {10},
  journal = {The Journal of Chemical Physics},
  publisher = {AIP Publishing},
  author = {Helmich-Paris,  Benjamin and de Souza,  Bernardo and Neese,  Frank and Izsák,  Róbert},
  year = {2021},
}

@article{Kollmar2021,
  title = {An efficient implementation of the NEVPT2 and CASPT2 methods avoiding higher-order density matrices},
  volume = {155},
  ISSN = {1089-7690},
  url = {http://dx.doi.org/10.1063/5.0072129},
  DOI = {10.1063/5.0072129},
  number = {23},
  journal = {The Journal of Chemical Physics},
  publisher = {AIP Publishing},
  author = {Kollmar,  Christian and Sivalingam,  Kantharuban and Guo,  Yang and Neese,  Frank},
  year = {2021},
  month = Dec 
}

@article{Neese2022,
  title = {The
                    <scp>SHARK</scp>
                    integral generation and digestion system},
  volume = {44},
  ISSN = {1096-987X},
  url = {http://dx.doi.org/10.1002/jcc.26942},
  DOI = {10.1002/jcc.26942},
  number = {3},
  journal = {Journal of Computational Chemistry},
  publisher = {Wiley},
  author = {Neese,  Frank},
  year = {2022},
  pages = {381–396}
}

@article{Giannozzi2009,
  title = {QUANTUM ESPRESSO: a modular and open-source software project for quantum simulations of materials},
  volume = {21},
  ISSN = {1361-648X},
  url = {http://dx.doi.org/10.1088/0953-8984/21/39/395502},
  DOI = {10.1088/0953-8984/21/39/395502},
  number = {39},
  journal = {Journal of Physics: Condensed Matter},
  publisher = {IOP Publishing},
  author = {Giannozzi,  Paolo and Baroni,  Stefano and Bonini,  Nicola and Calandra,  Matteo and Car,  Roberto and Cavazzoni,  Carlo and Ceresoli,  Davide and Chiarotti,  Guido L and Cococcioni,  Matteo and Dabo,  Ismaila and Dal Corso,  Andrea and de Gironcoli,  Stefano and Fabris,  Stefano and Fratesi,  Guido and Gebauer,  Ralph and Gerstmann,  Uwe and Gougoussis,  Christos and Kokalj,  Anton and Lazzeri,  Michele and Martin-Samos,  Layla and Marzari,  Nicola and Mauri,  Francesco and Mazzarello,  Riccardo and Paolini,  Stefano and Pasquarello,  Alfredo and Paulatto,  Lorenzo and Sbraccia,  Carlo and Scandolo,  Sandro and Sclauzero,  Gabriele and Seitsonen,  Ari P and Smogunov,  Alexander and Umari,  Paolo and Wentzcovitch,  Renata M},
  year = {2009},
  pages = {395502}
}

@article{Giannozzi2017,
  title = {Advanced capabilities for materials modelling with Q
                    <scp>uantum</scp>
                    ESPRESSO},
  volume = {29},
  ISSN = {1361-648X},
  url = {http://dx.doi.org/10.1088/1361-648X/aa8f79},
  DOI = {10.1088/1361-648x/aa8f79},
  number = {46},
  journal = {Journal of Physics: Condensed Matter},
  publisher = {IOP Publishing},
  author = {Giannozzi,  P and Andreussi,  O and Brumme,  T and Bunau,  O and Buongiorno Nardelli,  M and Calandra,  M and Car,  R and Cavazzoni,  C and Ceresoli,  D and Cococcioni,  M and Colonna,  N and Carnimeo,  I and Dal Corso,  A and de Gironcoli,  S and Delugas,  P and DiStasio,  R A and Ferretti,  A and Floris,  A and Fratesi,  G and Fugallo,  G and Gebauer,  R and Gerstmann,  U and Giustino,  F and Gorni,  T and Jia,  J and Kawamura,  M and Ko,  H-Y and Kokalj,  A and K\"{u}\c{c}\"{u}kbenli,  E and Lazzeri,  M and Marsili,  M and Marzari,  N and Mauri,  F and Nguyen,  N L and Nguyen,  H-V and Otero-de-la-Roza,  A and Paulatto,  L and Poncé,  S and Rocca,  D and Sabatini,  R and Santra,  B and Schlipf,  M and Seitsonen,  A P and Smogunov,  A and Timrov,  I and Thonhauser,  T and Umari,  P and Vast,  N and Wu,  X and Baroni,  S},
  year = {2017},
  month = Oct,
  pages = {465901}
}

@article{neu_single_2011,
	title = {Single photon emission from silicon-vacancy colour centres in chemical vapour deposition nano-diamonds on iridium},
	volume = {13},
	number = {2},
	journal = {New J. Phys.},
	author = {Neu, Elke and Steinmetz, David and Riedrich-Möller, Janine and Gsell, Stefan and Fischer, Martin and Schreck, Matthias and Becher, Christoph},
	year = {2011},
	pages = {025012},
}

@article{muller_optical_2014,
	title = {Optical signatures of silicon-vacancy spins in diamond},
	volume = {5},
	journal = {Nat. Commun.},
	author = {Müller, Tina and Hepp, Christian and Pingault, Benjamin and Neu, Elke and Gsell, Stefan and Schreck, Matthias and Sternschulte, Hadwig and Steinmüller-Nethl, Doris and Becher, Christoph and Atatüre, Mete},
	year = {2014},
	pages = {3328},
}

@article{iwasaki_germanium-vacancy_2015,
	title = {Germanium-{Vacancy} {Single} {Color} {Centers} in {Diamond}},
	volume = {5},
	number = {1},
	journal = {Scientific Reports},
	author = {Iwasaki, Takayuki and Ishibashi, Fumitaka and Miyamoto, Yoshiyuki and Doi, Yuki and Kobayashi, Satoshi and Miyazaki, Takehide and Tahara, Kosuke and Jahnke, Kay D. and Rogers, Lachlan J. and Naydenov, Boris and Jelezko, Fedor and Yamasaki, Satoshi and Nagamachi, Shinji and Inubushi, Toshiro and Mizuochi, Norikazu and Hatano, Mutsuko},
	year = {2015},
	pages = {12882},
}

@article{iwasaki_tin-vacancy_2017,
	title = {Tin-{Vacancy} {Quantum} {Emitters} in {Diamond}},
	volume = {119},
	number = {25},
	journal = {Phys. Rev. Lett.},
	author = {Iwasaki, Takayuki and Miyamoto, Yoshiyuki and Taniguchi, Takashi and Siyushev, Petr and Metsch, Mathias H. and Jelezko, Fedor and Hatano, Mutsuko},
	year = {2017},
	pages = {253601},
}

@article{trusheim_transform-limited_2020,
	title = {Transform-{Limited} {Photons} {From} a {Coherent} {Tin}-{Vacancy} {Spin} in {Diamond}},
	volume = {124},
	number = {2},
	journal = {Phys. Rev. Lett.},
	author = {Trusheim, Matthew E. and Pingault, Benjamin and Wan, Noel H. and Gündoğan, Mustafa and De Santis, Lorenzo and Debroux, Romain and Gangloff, Dorian and Purser, Carola and Chen, Kevin C. and Walsh, Michael and Rose, Joshua J. and Becker, Jonas N. and Lienhard, Benjamin and Bersin, Eric and Paradeisanos, Ioannis and Wang, Gang and Lyzwa, Dominika and Montblanch, Alejandro R-P. and Malladi, Girish and Bakhru, Hassaram and Ferrari, Andrea C. and Walmsley, Ian A. and Atatüre, Mete and Englund, Dirk},
	year = {2020},
	pages = {023602},
}

@article{wang_low-temperature_2021,
	title = {Low-{Temperature} {Spectroscopic} {Investigation} of {Lead}-{Vacancy} {Centers} in {Diamond} {Fabricated} by {High}-{Pressure} and {High}-{Temperature} {Treatment}},
	volume = {8},
	number = {10},
	journal = {ACS Photonics},
	author = {Wang, Peng and Taniguchi, Takashi and Miyamoto, Yoshiyuki and Hatano, Mutsuko and Iwasaki, Takayuki},
	year = {2021},
	pages = {2947},	
}

@article{wang_transform-limited_2024,
  title={Neutral silicon vacancy centers in undoped diamond via surface control},
  author={Zhang, Zi-Huai and Zuber, Josh A and Rodgers, Lila VH and Gui, Xin and Stevenson, Paul and Li, Minghao and Batzer, Marietta and Grimau Puigibert, Marcel li and Shields, Brendan J and Edmonds, Andrew M and others},
  journal={Phys. Rev. Lett.},
  volume={130},
  number={16},
  pages={166902},
  year={2023},
  publisher={APS}
}

@article{aharonovich_photophysics_2010,
	title = {Photophysics of chromium-related diamond single-photon emitters},
	journal = {Phys. Rev. A},
	volume = {81},
	pages = {043813},
	number = {4},

	author = {Aharonovich, I. and Castelletto, S. and Simpson, D. A. and Greentree, A. D. and Prawer, S.},
	date = {2010-04-13},
    year=2010
}

@article{hermans_qubit_2022,
	title = {Qubit teleportation between non-neighbouring nodes in a quantum network},
	volume = {605},
	number = {7911},
	journal = {Nature},
	author = {Hermans, S. L. N. and Pompili, M. and Beukers, H. K. C. and Baier, S. and Borregaard, J. and Hanson, R.},
	year = {2022},
	pages = {663},
}

@article{
pompili_realization_nodate,
author = {M. Pompili  and S. L. N. Hermans  and S. Baier  and H. K. C. Beukers  and P. C. Humphreys  and R. N. Schouten  and R. F. L. Vermeulen  and M. J. Tiggelman  and L. dos Santos Martins  and B. Dirkse  and S. Wehner  and R. Hanson },
title = {Realization of a multinode quantum network of remote solid-state qubits},
journal = {Science},
volume = {372},
number = {6539},
pages = {259-264},
year = {2021},
doi = {10.1126/science.abg1919},
eprint = {https://www.science.org/doi/pdf/10.1126/science.abg1919}}

@article{hensen_loophole-free_2015,
	title = {Loophole-free {Bell} inequality violation using electron spins separated by 1.3 kilometres},
	volume = {526},
	number = {7575},
	journal = {Nature},
	author = {Hensen, B. and Bernien, H. and Dréau, A. E. and Reiserer, A. and Kalb, N. and Blok, M. S. and Ruitenberg, J. and Vermeulen, R. F. L. and Schouten, R. N. and Abellán, C. and Amaya, W. and Pruneri, V. and Mitchell, M. W. and Markham, M. and Twitchen, D. J. and Elkouss, D. and Wehner, S. and Taminiau, T. H. and Hanson, R.},
	year = {2015},
	pages = {682},
}

@article{goss_density_2024,
	title = {Density functional theory study of {Al}, {Ga} and in impurities in diamond},
	volume = {142},
	issn = {09259635},
	url = {https://linkinghub.elsevier.com/retrieve/pii/S0925963524000244},
	doi = {10.1016/j.diamond.2024.110811},
	language = {en},
	urldate = {2025-09-17},
	journal = {Diam. Relat. Mater.},
	author = {Goss, J.P. and Lowery, Ruairi and Briddon, P.R. and Rayson, M.J.},
	month = feb,
	year = {2024},
	pages = {110811},
}

@misc{xiong_identifying_2025,
      title={Identifying high performance spectrally-stable quantum defects in diamond}, 
      author={Yihuang Xiong and Yizhi Zhu and Shay McBride and Sinéad M. Griffin and Geoffroy Hautier},
      year={2025},
      eprint={2504.11598},
      archivePrefix={arXiv},
      primaryClass={cond-mat.mtrl-sci},
      url={https://arxiv.org/abs/2504.11598}, 
}

@article{xiong_computationally_2024,
	title = {Computationally {Driven} {Discovery} of {T} {Center}-like {Quantum} {Defects} in {Silicon}},
	volume = {146},
	number = {44},
	journal = {J. Am. Chem. Soc.},
	author = {Xiong, Yihuang and Zheng, Jiongzhi and McBride, Shay and Zhang, Xueyue and Griffin, Sinéad M. and Hautier, Geoffroy},
	year = {2024},
	pages = {30046},
}

@article{thiering_eg_2019,
	title = {The (eg ⊗ eu) ⊗ {Eg} product {Jahn}–{Teller} effect in the neutral group-{IV} vacancy quantum bits in diamond},
	volume = {5},
	issn = {2057-3960},
	url = {https://www.nature.com/articles/s41524-019-0158-3},
	doi = {10.1038/s41524-019-0158-3},
	language = {en},
	number = {1},
	urldate = {2025-02-13},
	journal = {npj Comput. Mater.},
	author = {Thiering, Gergő and Gali, Adam},
	month = feb,
	year = {2019},
	pages = {18},
}

@article{
xiong_high-throughput_2023,
author = {Yihuang Xiong  and Céline Bourgois  and Natalya Sheremetyeva  and Wei Chen  and Diana Dahliah  and Hanbin Song  and Jiongzhi Zheng  and Sinéad M. Griffin  and Alp Sipahigil  and Geoffroy Hautier },
title = {High-throughput identification of spin-photon interfaces in silicon},
journal = {Sci. Adv.},
volume = {9},
number = {40},
pages = {eadh8617},
year = {2023},
doi = {10.1126/sciadv.adh8617},
URL = {https://www.science.org/doi/abs/10.1126/sciadv.adh8617},
eprint = {https://www.science.org/doi/pdf/10.1126/sciadv.adh8617}}

@article{qiu_origin_2024,
	title = {Origin and designing of large ground-state zero-field splitting of color centers in diamond},
	volume = {110},
	number = {2},
	urldate = {2024-07-12},
	journal = {Phys. Rev. B},
	author = {Qiu, Chen and Liang, Han-Pu and Deng, Hui-Xiong and Wei, Su-Huai},
	year = {2024},
	pages = {024104},
}

@article{thiel_probing_2019,
	title = {Probing magnetism in {2D} materials at the nanoscale with single-spin microscopy},
	volume = {364},
	issn = {0036-8075, 1095-9203},
	url = {https://www.science.org/doi/10.1126/science.aav6926},
	doi = {10.1126/science.aav6926},
	language = {en},
	number = {6444},
	urldate = {2024-04-22},
	journal = {Science},
	author = {Thiel, L. and Wang, Z. and Tschudin, M. A. and Rohner, D. and Gutiérrez-Lezama, I. and Ubrig, N. and Gibertini, M. and Giannini, E. and Morpurgo, A. F. and Maletinsky, P.},
	month = jun,
	year = {2019},
	pages = {973--976},
}

@article{rose_observation_2018,
	title = {Observation of an environmentally insensitive solid-state spin defect in diamond},
	volume = {361},
	issn = {0036-8075, 1095-9203},
	url = {https://www.science.org/doi/10.1126/science.aao0290},
	doi = {10.1126/science.aao0290},
	language = {en},
	number = {6397},
	urldate = {2024-04-14},
	journal = {Science},
	author = {Rose, Brendon C. and Huang, Ding and Zhang, Zi-Huai and Stevenson, Paul and Tyryshkin, Alexei M. and Sangtawesin, Sorawis and Srinivasan, Srikanth and Loudin, Lorne and Markham, Matthew L. and Edmonds, Andrew M. and Twitchen, Daniel J. and Lyon, Stephen A. and De Leon, Nathalie P.},
	month = jul,
	year = {2018},
	pages = {60--63},
}

@article{henshaw_nanoscale_2022,
	title = {Nanoscale solid-state nuclear quadrupole resonance spectroscopy using depth-optimized nitrogen-vacancy ensembles in diamond},
	volume = {120},
	issn = {0003-6951, 1077-3118},
	url = {https://pubs.aip.org/apl/article/120/17/174002/2833616/Nanoscale-solid-state-nuclear-quadrupole-resonance},
	doi = {10.1063/5.0083774},
	language = {en},
	number = {17},
	urldate = {2024-04-14},
	journal = {Appl. Phys. Lett.},
	author = {Henshaw, Jacob and Kehayias, Pauli and Saleh Ziabari, Maziar and Titze, Michael and Morissette, Erin and Watanabe, Kenji and Taniguchi, Takashi and Li, J. I. A. and Acosta, Victor M. and Bielejec, Edward S. and Lilly, Michael P. and Mounce, Andrew M.},
	month = apr,
	year = {2022},
	pages = {174002},
}

@article{Awschalom2018, 
title={Quantum technologies with optically interfaced solid-state spins},
volume={12}, 
number={9}, 
journal={Nat. Photon.}, 
author={Awschalom, David D. and Hanson, Ronald and Wrachtrup, Jörg and Zhou, Brian B.}, 
year={2018}, 
pages={516}}

@article{Gordon_2013, 
title={Quantum computing with defects}, 
volume={38}, 
number={10}, 
journal={MRS Bulletin}, 
author={Gordon, Luke and Weber, Justin R. and Varley, Joel B. and Janotti, Anderson and Awschalom, David D. and Van De Walle, Chris G.}, 
year={2013}, 
pages={802}}

@article{Wolfowicz_2021, 
title={Quantum guidelines for solid-state spin defects}, 
volume={6}, 
number={10}, 
journal={Nat. Rev. Mater.}, 
author={Wolfowicz, Gary and Heremans, F. Joseph and Anderson, Christopher P. and Kanai, Shun and Seo, Hosung and Gali, Adam and Galli, Giulia and Awschalom, David D.}, 
year={2021},
pages={906}}

@article{harris_group-iii_2020,
	title = {Group-{III} quantum defects in diamond are stable spin-1 color centers},
	volume = {102},
	issn = {2469-9950, 2469-9969},
	url = {https://link.aps.org/doi/10.1103/PhysRevB.102.195206},
	doi = {10.1103/PhysRevB.102.195206},
	pages = {195206},
	number = {19},
	journaltitle = {Phys. Rev. B},
	journal = {Phys. Rev. B},
	shortjournal = {Phys. Rev. B},
	author = {Harris, Isaac and Ciccarino, Christopher J. and Flick, Johannes and Englund, Dirk R. and Narang, Prineha},
	urldate = {2024-07-30},
	date = {2020-11-24},
	langid = {english},
    year=2020,
}

@article{dhaenens-johansson_optical_2011,
	title = {Optical properties of the neutral silicon split-vacancy center in diamond},
	author = {D'Haenens-Johansson, U. F. S. and Edmonds, A. M. and Green, B. L. and Newton, M. E. and Davies, G. and Martineau, P. M. and Khan, R. U. A. and Twitchen, D. J.},
	journal = {Phys. Rev. B},
	urldate = {2026-01-08},
	volume = {84},
	rights = {http://link.aps.org/licenses/aps-default-license},
	issn = {1098-0121, 1550-235X},
	url = {https://link.aps.org/doi/10.1103/PhysRevB.84.245208},
	doi = {10.1103/PhysRevB.84.245208},
	pages = {245208},
	number = {24},
	date = {2011-12-21},
	langid = {english},
    year=2011,
}

@article{green_electronic_2019,
	title = {Electronic structure of the neutral silicon-vacancy center in diamond},
	author = {Green, B. L. and Doherty, M. W. and Nako, E. and Manson, N. B. and D'Haenens-Johansson, U. F. S. and Williams, S. D. and Twitchen, D. J. and Newton, M. E.},
	journal = {Phys. Rev. B},
	urldate = {2026-01-08},
	volume = {99},
	issn = {2469-9950, 2469-9969},
	url = {https://link.aps.org/doi/10.1103/PhysRevB.99.161112},
	doi = {10.1103/PhysRevB.99.161112},
	pages = {161112},
	number = {16},
	date = {2019-04-17},
	langid = {english},
    year=2019,
}

@article{zhang_neutral_2023,
	title = {Neutral silicon vacancy centers in undoped diamond via surface control},
	volume = {130},
	issn = {0031-9007, 1079-7114},
	url = {http://arxiv.org/abs/2206.13698},
	doi = {10.1103/PhysRevLett.130.166902},
	pages = {166902},
	number = {16},
	journaltitle = {Phys. Rev. Lett.},
	journal = {Phys. Rev. Lett.},
	shortjournal = {Phys. Rev. Lett.},
	author = {Zhang, Zi-Huai and Zuber, Josh A. and Rodgers, Lila V. H. and Gui, Xin and Stevenson, Paul and Li, Minghao and Batzer, Marietta and Grimau, Marcel li and Shields, Brendan and Edmonds, Andrew M. and Palmer, Nicola and Markham, Matthew L. and Cava, Robert J. and Maletinsky, Patrick and de Leon, Nathalie P.},
	urldate = {2024-04-14},
	date = {2023-04-21},
	langid = {english},
	eprinttype = {arxiv},
	eprint = {2206.13698 [physics, physics:quant-ph]},
    year=2023
}

@article{maze_nanoscale_2008,
	title = {Nanoscale magnetic sensing with an individual electronic spin in diamond},
	volume = {455},
	copyright = {2008 Macmillan Publishers Limited. All rights reserved},
	issn = {1476-4687},
	url = {https://www.nature.com/articles/nature07279},
	doi = {10.1038/nature07279},
	language = {en},
	number = {7213},
	urldate = {2025-10-28},
	journal = {Nature},
	author = {Maze, J. R. and Stanwix, P. L. and Hodges, J. S. and Hong, S. and Taylor, J. M. and Cappellaro, P. and Jiang, L. and Dutt, M. V. Gurudev and Togan, E. and Zibrov, A. S. and Yacoby, A. and Walsworth, R. L. and Lukin, M. D.},
	month = oct,
	year = {2008},
	pages = {644--647},
}

@article{doherty_nitrogen-vacancy_2013,
	series = {The nitrogen-vacancy colour centre in diamond},
	title = {The nitrogen-vacancy colour centre in diamond},
	volume = {528},
	issn = {0370-1573},
	url = {https://www.sciencedirect.com/science/article/pii/S0370157313000562},
	doi = {10.1016/j.physrep.2013.02.001},
	number = {1},
	urldate = {2025-10-28},
	journal = {Phys. Rep.},
	author = {Doherty, Marcus W. and Manson, Neil B. and Delaney, Paul and Jelezko, Fedor and Wrachtrup, Jörg and Hollenberg, Lloyd C. L.},
	month = jul,
	year = {2013},
	pages = {1--45},
}

@article{rondin_magnetometry_2014,
	title = {Magnetometry with nitrogen-vacancy defects in diamond},
	volume = {77},
	issn = {0034-4885},
	url = {https://doi.org/10.1088/0034-4885/77/5/056503},
	doi = {10.1088/0034-4885/77/5/056503},
	language = {en},
	number = {5},
	urldate = {2025-10-28},
	journal = {Rep. Prog. Phys.},
	author = {Rondin, L and Tetienne, J-P and Hingant, T and Roch, J-F and Maletinsky, P and Jacques, V},
	month = may,
	year = {2014},
	pages = {056503},
}

@article{abe_tutorial_2018,
	title = {Tutorial: {Magnetic} resonance with nitrogen-vacancy centers in diamond—microwave engineering, materials science, and magnetometry},
	volume = {123},
	issn = {0021-8979},
	shorttitle = {Tutorial},
	url = {https://doi.org/10.1063/1.5011231},
	doi = {10.1063/1.5011231},
	number = {16},
	urldate = {2025-10-28},
	journal = {J. Appl. Phys.},
	author = {Abe, Eisuke and Sasaki, Kento},
	month = mar,
	year = {2018},
	pages = {161101},
}

@article{levine_principles_2019,
	title = {Principles and techniques of the quantum diamond microscope},
	volume = {8},
	copyright = {De Gruyter expressly reserves the right to use all content for commercial text and data mining within the meaning of Section 44b of the German Copyright Act.},
	issn = {2192-8614},
	url = {https://www.degruyterbrill.com/document/doi/10.1515/nanoph-2019-0209/html?lang=en&srsltid=AfmBOopkejLF1g3ahMekrYkQDOLYB1kHx3OlbuFh57ikrmvG4cIi937o},
	doi = {10.1515/nanoph-2019-0209},
	language = {en},
	number = {11},
	urldate = {2025-10-28},
	journal = {Nanophotonics},
	author = {Levine, Edlyn V. and Turner, Matthew J. and Kehayias, Pauli and Hart, Connor A. and Langellier, Nicholas and Trubko, Raisa and Glenn, David R. and Fu, Roger R. and Walsworth, Ronald L.},
	month = nov,
	year = {2019},
	pages = {1945--1973},
}

@article{barry_sensitivity_2020,
	title = {Sensitivity optimization for {NV}-diamond magnetometry},
	volume = {92},
	issn = {0034-6861, 1539-0756},
	url = {https://link.aps.org/doi/10.1103/RevModPhys.92.015004},
	doi = {10.1103/RevModPhys.92.015004},
	language = {en},
	number = {1},
	urldate = {2025-10-14},
	journal = {Rev. Mod. Phys.},
	author = {Barry, John F. and Schloss, Jennifer M. and Bauch, Erik and Turner, Matthew J. and Hart, Connor A. and Pham, Linh M. and Walsworth, Ronald L.},
	month = mar,
	year = {2020},
	pages = {015004},
}

@article{xu_recent_2023,
	title = {Recent advances on applications of {NV}$^{-}$ magnetometry in condensed matter physics},
	volume = {11},
	copyright = {© 2023 Chinese Laser Press},
	issn = {2327-9125},
	url = {https://opg.optica.org/prj/abstract.cfm?uri=prj-11-3-393},
	doi = {10.1364/PRJ.471266},
	language = {EN},
	number = {3},
	urldate = {2025-10-28},
	journal = {Photon. Res., PRJ},
	author = {Xu, Ying and Zhang, Weiye and Tian, Chuanshan},
	month = mar,
	year = {2023},
	pages = {393--412},
}

@article{Pezzagna_NV_quantum_computer,
    author = {Pezzagna, Sébastien and Meijer, Jan},
    title = {Quantum computer based on color centers in diamond},
    journal = {Appl. Phys. Rev.},
    volume = {8},
    number = {1},
    pages = {011308},
    year = {2021},
    month = {02},
    issn = {1931-9401},
    doi = {10.1063/5.0007444},
    url = {https://doi.org/10.1063/5.0007444},
    eprint = {https://pubs.aip.org/aip/apr/article-pdf/doi/10.1063/5.0007444/19740523/011308_1_online.pdf},
}

@article{thomas_substitutional_2024,
	author = {Thomas, John C. and Chen, Wei and Xiong, Yihuang and Barker, Bradford A. and Zhou, Junze and Chen, Weiru and Rossi, Antonio and Kelly, Nolan and Yu, Zhuohang and Zhou, Da and Kumari, Shalini and Barnard, Edward S. and Robinson, Joshua A. and Terrones, Mauricio and Schwartzberg, Adam and Ogletree, D. Frank and Rotenberg, Eli and Noack, Marcus M. and Griffin, Sinéad and Raja, Archana and Strubbe, David A. and Rignanese, Gian-Marco and Weber-Bargioni, Alexander and Hautier, Geoffroy},
	title = {A substitutional quantum defect in {WS}2 discovered by high-throughput computational screening and fabricated by site-selective {STM} manipulation},
    journal = {Nat Commun},
	volume = {15},
	issn = {2041-1723},
	url = {https://www.nature.com/articles/s41467-024-47876-3},
	doi = {10.1038/s41467-024-47876-3},
	pages = {3556},
	number = {1},
	urldate = {2026-01-21},
	date = {2024-04-26},
	langid = {english},
    year=2024,
}

@article{Sajid_2D_High_throughput,
author = {Ali, Sajid and Nilsson, Fredrik Andreas and Manti, Simone and Bertoldo, Fabian and Mortensen, Jens Jørgen and Thygesen, Kristian Sommer},
title = {High-Throughput Search for Triplet Point Defects with Narrow Emission Lines in 2D Materials},
journal = {ACS Nano},
volume = {17},
number = {21},
pages = {21105-21115},
year = {2023},
doi = {10.1021/acsnano.3c04774},
URL = {https://doi.org/10.1021/acsnano.3c04774},
eprint = {https://doi.org/10.1021/acsnano.3c04774}

}

@misc{riendeau_quantum_2023,
	author = {Riendeau, Emmeline G. and Basso, Luca and Mah, Jasmine J. and Cong, Rong and Sadi, {MA} and Henshaw, Jacob and Azizur-Rahman, {KM} and Jones, Aulden and Joshi, Gajadhar and Lilly, Michael P. and Mounce, Andrew A.},
	title = {Quantum Instrumentation Control Kit -- Defect Arbitrary Waveform Generator ({QICK}-{DAWG}): A Quantum Sensing Control Framework for Quantum Defects},
	publisher = {{arXiv}},
	rights = {Creative Commons Attribution 4.0 International},
	url = {https://arxiv.org/abs/2311.18253},
	doi = {10.48550/ARXIV.2311.18253},
	urldate = {2026-02-04},
	date = {2023},
}

@article{garciaarellano_photoinduced_2024,
title = {Photo-Induced Charge State Dynamics of the Neutral and Negatively Charged Silicon Vacancy Centers in Room-Temperature Diamond},
author = {Garcia-Arellano, G. and López-Morales, G. I. and Manson, N. B. and Flick, J. and Wood, A. A. and Meriles, C. A.},
journal = {Adv. Sci.},
volume = {11},
number = {22},
pages = {2308814},
doi = {https://doi.org/10.1002/advs.202308814},
url = {https://advanced.onlinelibrary.wiley.com/doi/abs/10.1002/advs.202308814},
eprint = {https://advanced.onlinelibrary.wiley.com/doi/pdf/10.1002/advs.202308814},
year = {2024}
}

@article{NV_sensitivity_Barry,
  title = {Sensitivity optimization for NV-diamond magnetometry},
  author = {Barry, John F. and Schloss, Jennifer M. and Bauch, Erik and Turner, Matthew J. and Hart, Connor A. and Pham, Linh M. and Walsworth, Ronald L.},
  journal = {Rev. Mod. Phys.},
  volume = {92},
  issue = {1},
  pages = {015004},
  numpages = {68},
  year = {2020},
  month = {Mar},
  publisher = {American Physical Society},
  doi = {10.1103/RevModPhys.92.015004},
  url = {https://link.aps.org/doi/10.1103/RevModPhys.92.015004}
}

@article{
temp_sens_Awschalom,
author = {David M. Toyli  and Charles F. de las Casas  and David J. Christle  and Viatcheslav V. Dobrovitski  and David D. Awschalom },
title = {Fluorescence thermometry enhanced by the quantum coherence of single spins in diamond},
journal = {Proc. Natl. Acad. Sci. U.S.A.},
volume = {110},
number = {21},
pages = {8417-8421},
year = {2013},
doi = {10.1073/pnas.1306825110},
URL = {https://www.pnas.org/doi/abs/10.1073/pnas.1306825110},
eprint = {https://www.pnas.org/doi/pdf/10.1073/pnas.1306825110}}

@article{temp_sens_Gali,
  title = {Physically motivated analytical expression for the temperature dependence of the zero-field splitting of the nitrogen-vacancy center in diamond},
  author = {Cambria, M. C. and Thiering, G. and Norambuena, A. and Dinani, H. T. and Gardill, A. and Kemeny, I. and Lordi, V. and Gali, \'A. and Maze, J. R. and Kolkowitz, S.},
  journal = {Phys. Rev. B},
  volume = {108},
  issue = {18},
  pages = {L180102},
  numpages = {7},
  year = {2023},
  month = {Nov},
  publisher = {American Physical Society},
  doi = {10.1103/PhysRevB.108.L180102},
  url = {https://link.aps.org/doi/10.1103/PhysRevB.108.L180102}
}

@article{kresse1996PRB,
  title={Efficient iterative schemes for ab initio total-energy calculations using a plane-wave basis set},
  author={Kresse, Georg and Furthm{\"u}ller, J{\"u}rgen},
  journal={Phys. Rev. B},
  volume={54},
  number={16},
  pages={11169},
  year={1996},
  publisher={APS}
}

@article{blochl1994PRB,
  author = {Bl{\"o}chl, Peter E.},
  title = {Projector augmented-wave method},
  journal = {Phys. Rev. B},
  volume = {50},
  pages = {17953--17979},
  year = {1994},
  doi = {10.1103/PhysRevB.50.17953}
}

@article{kresse1999PRB,
  title={From ultrasoft pseudopotentials to the projector augmented-wave method},
  author={Kresse, Georg and Joubert, Daniel},
  journal={Phys. Rev. B},
  volume={59},
  number={3},
  pages={1758},
  year={1999},
  publisher={APS}
}

@article{kresse1996CMS,
  title={Efficiency of ab-initio total energy calculations for metals and semiconductors using a plane-wave basis set},
  author={Kresse, Georg and Furthm{\"u}ller, J{\"u}rgen},
  journal={Comput. Mater. Sci.},
  volume={6},
  number={1},
  pages={15--50},
  year={1996},
  publisher={Elsevier}
}

@article{kresse1993PRB,
  title={Ab initio molecular dynamics for liquid metals},
  author={Kresse, Georg and Hafner, J{\"u}rgen},
  journal={Phys. Rev. B},
  volume={47},
  number={1},
  pages={558},
  year={1993},
  publisher={APS}
}

@article{kresse1994PRB,
  title={Ab initio molecular-dynamics simulation of the liquid-metal--amorphous-semiconductor transition in germanium},
  author={Kresse, Georg and Hafner, J{\"u}rgen},
  journal={Phys. Rev. B},
  volume={49},
  number={20},
  pages={14251},
  year={1994},
  publisher={APS}
}

@article{kresse1994JPCM,
  title={Norm-conserving and ultrasoft pseudopotentials for first-row and transition elements},
  author={Kresse, Georg and Hafner, Jurgen},
  journal={J. Phys. Condens. Matter},
  volume={6},
  number={40},
  pages={8245},
  year={1994},
  publisher={IOP Publishing}
}

@article{heyd2003JCP,
  author = {Heyd, Jochen and Scuseria, Gustavo E. and Ernzerhof, Matthias},
  title = {Hybrid functionals based on a screened Coulomb potential},
  journal = {J. Chem. Phys.},
  volume = {118},
  pages = {8207--8215},
  year = {2003},
  doi = {10.1063/1.1564060}
}

@article{heyd2006JCP,
  author = {Heyd, Jochen and Scuseria, Gustavo E. and Ernzerhof, Matthias},
  title = {Erratum: Hybrid functionals based on a screened Coulomb potential},
  journal = {J. Chem. Phys.},
  volume = {124},
  pages = {219906},
  year = {2006},
  doi = {10.1063/1.2204597}
}

@article{marsman2010JCP,
  author = {Marsman, Martijn and Natu, Vikram and Harda, T. and Kresse, Georg},
  title = {Calculating zero-field splitting parameters using the projector augmented wave method},
  journal = {J. Chem. Phys.},
  volume = {152},
  pages = {104102},
  year = {2020},
  doi = {10.1063/1.5142395}
}

@article{
doi:10.1073/pnas.2422542122,
author = {Uri Zvi  and Denis R. Candido  and Adam M. Weiss  and Aidan R. Jones  and Lingjie Chen  and Iryna Golovina  and Xiaofei Yu  and Stella Wang  and Dmitri V. Talapin  and Michael E. Flatté  and Aaron P. Esser-Kahn  and Peter C. Maurer },
title = {Engineering spin coherence in core-shell diamond nanocrystals},
journal = {Proc. Natl. Acad. Sci. U.S.A.},
volume = {122},
number = {21},
pages = {e2422542122},
year = {2025},
doi = {10.1073/pnas.2422542122},
URL = {https://www.pnas.org/doi/abs/10.1073/pnas.2422542122},
eprint = {https://www.pnas.org/doi/pdf/10.1073/pnas.2422542122}}

\bibliographystyle{Science}

\newpage

\setcounter{figure}{0}
\renewcommand{\thefigure}{S\arabic{figure}}

\setcounter{table}{0}
\renewcommand{\thetable}{S\arabic{table}}

\section{Supplementary Information}

\begin{figure}[h]
    \centering
    \includegraphics[width=1\linewidth]{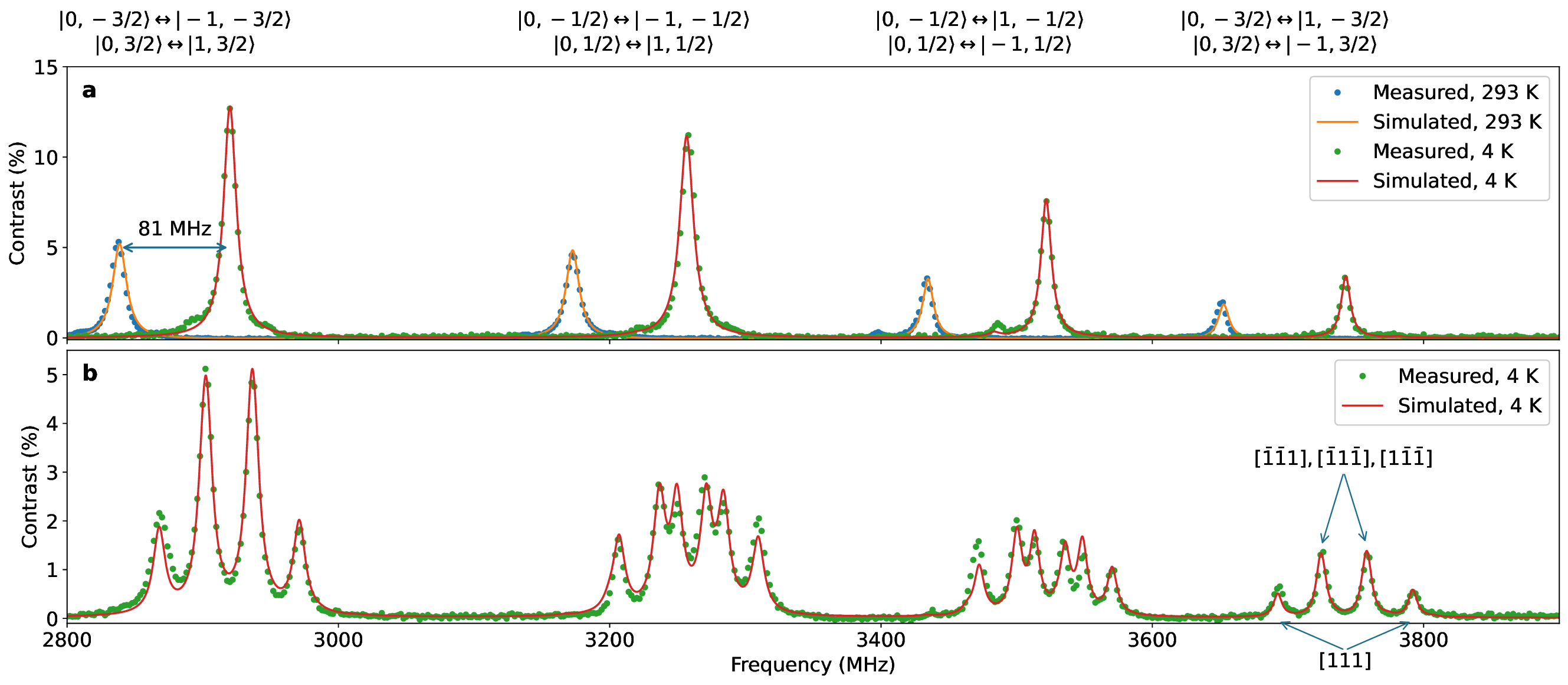}
    \caption{Comparison between measured ODMR at $T = 293$ K (room temperature) and $T = 4$ K. At $T = $4 K, the measured ODMR peaks shift by 83 MHz to higher frequency relative to room temperature.}
    \label{4K ODMR}
\end{figure}

\begin{table}[h]
\begin{tabular}{|c|c|c|c|c|}
\hline
Property & 293 K, Zero Field & 293 K, In-field & 4 K, Zero Field & 4 K, In-field \\\hline
$D$ (MHz) & 3192 $\pm$ 0.06 & 3190 $\pm$ 0.32 & 3277 $\pm$ 0.02 & 3274 $\pm$ 1.79 \\
$E$ (MHz) & -3.56 $\pm$ 2e-6  & -3.46 $\pm$ 3.42 & -3.56 $\pm$ 3e-5 & -3.66 $\pm$ 12.85 \\
$A_{iso}$ (MHz) & -273 $\pm$ 0.01  & -273.3 $\pm$ 0.49 & -276 $\pm$ 0.04 & -281 $\pm$ 2.7 \\
$A_{ax}$ (MHz) & 6.88 $\pm$ 3.4e-4 & 6.98 $\pm$ 0.29 & 6.88 $\pm$ 4.7e-4 & 9.78 $\pm$ 1.4 \\\hline
\end{tabular}

\caption{Fit values from the ODMR spectra of GaV$^-$ at 4 K and 293 K}\label{GaV Fit Values}
\end{table}

\end{document}